\documentclass[structabstract]{aa}  
\usepackage{graphicx}
\usepackage{txfonts}
\usepackage{natbib}

\newcommand{\fig}[1]{Fig.~\ref{#1}}
\newcommand{\sect}[1]{Sect.~\ref{#1}}

\newcommand{\referee}[1]{#1}

\begin{document}

   \title{Accounting for the {XRT} early steep decay \\ in models of the prompt GRB emission}

\author{
   	R. Hasco\"et, 
         F. Daigne \thanks{Institut Universitaire de France}
          \and
          R. Mochkovitch
          }

\institute{UPMC-CNRS, UMR7095, Institut d'Astrophysique de Paris, F-75014, Paris, France \\
               \email{[hascoet;daigne;mochko]@iap.fr}}

   \date{Received 4 April 2012 / Accepted 19 May 2012}
 
 \authorrunning{Hasco\"et, Daigne \& Mochkovitch}  
 \titlerunning{%Accounting for 
 The {XRT} early steep decay and prompt GRB models}
% in models of the prompt GRB emission}
 
  \abstract
  % context  
   {\referee{The} \textit{Swift}-{XRT} observations of the early X-ray afterglow of GRBs show that it usually begins with 
a steep decay phase. }
  % aims 
   {A possible origin of this early steep decay is the high latitude emission that subsists  
when the on-axis emission of the last dissipating regions in the relativistic outflow has been switched-off. 
%We want to explore the consequences of this interpretation for various models of the prompt emission 
We 
%want to determine 
\referee{wish to establish
which of 
%the 
various} 
models of the prompt emission are compatible with this interpretation. 
}
  % methods 
   {We successively consider internal shocks, photospheric emission\referee{,} and magnetic reconnection and obtain 
the typical decay \referee{timescale} at the end of the prompt phase in each case. }
  % results 
   {Only internal shocks naturally predict a decay \referee{timescale} comparable to the burst duration, as required
 to explain {XRT} observations 
 %by 
 \referee{in terms of}
 high latitude emission. 
%It
\referee{The decay timescale of the high latitude emission}
 is much too short in photospheric models and the observed decay must then correspond
to an effective and generic behavior of the central engine. 
Reconnection models require some \textit{ad hoc} assumptions to 
agree with the data, which will have to be validated when a better description of the reconnection process becomes available.
  }
  % conclusions  
 {} 


   \maketitle

\section{Introduction}

Thanks to its precise localization capabilities followed by rapid slewing, the \textit{Swift} satellite \citep{gehrels_2004} can quickly -- typically in less 
than two minutes after a Gamma-ray burst (GRB) trigger -- \referee{repoint} its X-Ray Telescope ({XRT}, \citealt{burrows_2005}) toward the source. This achievement has %filled 
\referee{helped to fill}
the gap \referee{in observations}
between the prompt and late afterglow emissions and revealed the complexity of the early X-ray afterglow 
\citep{tagliaferri_2005, nousek_2006, obrien_2006}. Despite this complexity, the early steep decay 
that ends the prompt emission appears to be a generic behavior, common to most long GRBs. 
During this phase\referee{,} the flux decays with a temporal index $\alpha \simeq 3-5$ (with $F_{\nu} \propto t^{-\alpha}$).
It lasts for a typical duration $t_{\rm ESD}\sim 10^{2}-10^{3}$ s and is usually followed by the shallow or normal decay phases. 

The rapid gamma-ray light curve variability suggests that prompt emission has to be produced by internal mechanisms \citep{sari_1997},
%The prompt emission has to be produced by internal mechanisms (because of the strong gamma-ray light curve variability, \citealt{sari_1997}) 
while the later shallow and normal decay phases are usually attributed to %the 
\referee{deceleration} by the external medium (e.g. \citealt{meszaros_1997, sari_1998, rees_1998}).
As the backward extrapolation of the early steep decay \referee{connects reasonably} to the end of the prompt emission, it is often 
interpreted as its tail. It has moreover been shown that it is too steep to 
result from the interaction with the external medium (see e.g. \citealt{lazzati_2006}). 

One of the most discussed and natural scenario to account for the early steep decay had been described by \citet{kumar_2000} before 
the launch of \textit{Swift}. It explains this phase by the residual off-axis emission -- or high latitude emission -- 
that becomes visible when the on-axis prompt activity switches off. This scenario gives simple predictions and several 
studies have been dedicated to check whether it agrees with observations 
(see e.g. \citealt{liang_2006, butler_2007, zhang_2007, qin_2008, barniol_2009}). 
%The recent work by 
\referee{
%based on
On the basis of a realistic multiple pulse fitting approach proposed by \citet{genet_2009},
\citet{willingale_2010}  
%has 
confirmed} that the observed temporal slope of the early steep decay can be well explained in this context\referee{,} while 
the accompanying spectral softening is (at least qualitatively) also reproduced.

In this letter, we first discuss in \sect{section_constraints} some constraints implied by the high latitude scenario on the typical 
radius $R_{\gamma}$ where the prompt emission ends. We then 
investigate in \sect{section_comparison} if they are consistent with the predictions of the most discussed models for the prompt phase: 
internal shocks, \referee{Comptonized} photospheric emission\referee{,} and magnetic reconnection.
We summarize our results in \sect{section_conclusion}, which 
%is 
also 
\referee{presents our conclusions.}
%the 
%conclusion.  
\vspace{-2ex}
\section{Constraints on the prompt emission radius in the high latitude scenario}
\label{section_constraints}

The high latitude interpretation of the early steep decay constrains the value of $R_{\gamma}$ in two different ways:
\\
{\it (i)} The maximum possible duration of the high latitude emission
depends on  the opening angle $\theta$ of the jet and 
 is given by
\begin{equation}
t_{\rm HLE}=\frac{R_{\gamma}\,\theta^2}{2\,c}\ .%,
%{R_{\gamma}\,\theta^2\over 2\,c}\ ,
\end{equation}
%where $\theta$ is the opening angle of the jet.
 It has to be larger than the total duration of the early steep decay, 
$t_{\rm ESD}$,
which leads to \citep{lyutikov_2006,lazzati_2006}
\begin{equation}
\label{eqn_contrainte1}
R_{\gamma} \ga 6 \ 10^{14} \,({t_{\rm ESD}/100 \ {\rm s}})\;({\theta/ 0.1\ {\rm rd}})^{-1}
\rm \ cm\ .
\end{equation}

\noindent {\it (ii)} In a logarithmic plot 
%with 
\referee{where}
the the burst trigger 
%as
\referee{is} 
the origin of time, the high latitude emission has to behave 
as a power-law that smoothly connects 
to the end of the prompt phase with an initial decay index $\alpha\sim 3 - 5$. 

After the prompt emission ends at $R_{\gamma}$\referee{,} 
the high latitude (bolometric) flux received by the observer 
has been derived by several authors (see e.g. \citealt{kumar_2000, beloborodov_2011}). It takes the form
%takes the form \citep{kumar_2000} 
%
\begin{equation}
\label{flux_bol_flash}
F_{\rm bol}\propto 
\left(1+{t-t_{\rm burst}\over \tau_{\rm HLE}}\right)^{-3}\ ,
%{1\over \left(1+{t-t_{\rm burst}\over \tau_{\rm HLE}}\right)^3}\ ,
\end{equation}  
with 
\begin{equation}
\tau_{\rm HLE}\simeq R_{\gamma} / 2 \Gamma^{2} c\ ,
\end{equation}
where $\Gamma$ is the Lorentz factor of the emitting shell. In Eq. \ref{flux_bol_flash},
$t_{\rm burst}$ is the arrival time of the last on-axis photons, emitted at $R_{\gamma}$, and therefore corresponds 
to the duration of the prompt phase.
The description of the flux given by Eq. \ref{flux_bol_flash} remains valid as long as a new emission component does not emerge leading to the shallow/normal decay phase of the afterglow.
From Eq. \ref{flux_bol_flash}, the initial temporal decay index of the high latitude emission equals
\begin{equation}
\label{eqn_alpha}
\alpha=3\,{t_{\rm burst}\over \tau_{\rm HLE}}\ ,
\end{equation}   
which 
%imposes 
\referee{implies}
that $\tau_{\rm HLE} \simeq t_{\rm burst}$ to reproduce the observations, yielding a new constraint on $R_{\gamma}$
\begin{equation}
\label{eqn_contrainte2}
R_{\gamma} \simeq R_*=6\,10^{15}\,({t_{\rm burst}/ 10\ {\rm s}})\,({\Gamma/100})^2 \ \mathrm{cm} .
\end{equation}
If conversely $\tau_{\rm HLE} \ll t_{\rm burst}$ (or $R_{\gamma} \ll R_*$), 
the high latitude flux will experience a huge drop (by a factor of about $\left(t_{\rm burst}/\tau_{\rm HLE}\right)^3$ 
%right 
\referee{immediately}
after the prompt  
phase before recovering a slope $\alpha\sim 3$ only after several $t_{\rm burst}$.

It should be noted however that Eq. \ref{flux_bol_flash} and the discussion that follows 
%suppose 
\referee{assume}
that the emitting surface at $R_{\gamma}$ has a spherical shape 
and that the emission is isotropic in the comoving frame of 
the emitting material. This is not necessarily true (see \sect{section_comparison} below) so that the constraints given by 
Eqs.~\ref{eqn_contrainte1}-\ref{eqn_contrainte2} are only estimates that might be relaxed by a factor 
of a few, but certainly not by several order of magnitudes. 
%Note also
\referee{We also note} that the two constraints are compatible.

\section{Comparison with the emission radii predicted in different GRB models}
\label{section_comparison}

\subsection{Internal shocks}
\label{subsect_internal_shocks}

\begin{figure*}
\begin{center}
\begin{tabular}{cc}
\includegraphics[scale=0.275]{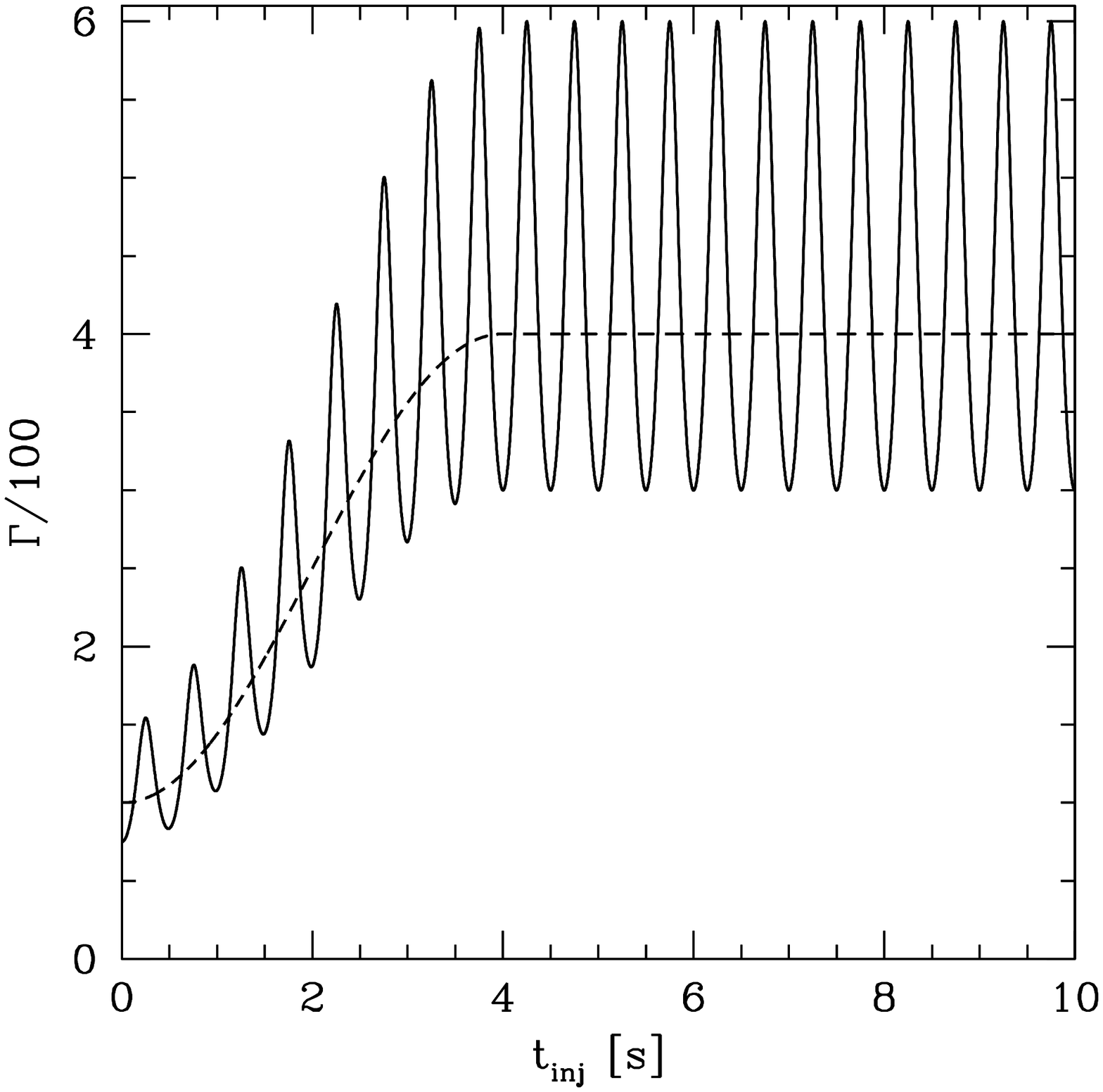} & \includegraphics[scale=0.275]{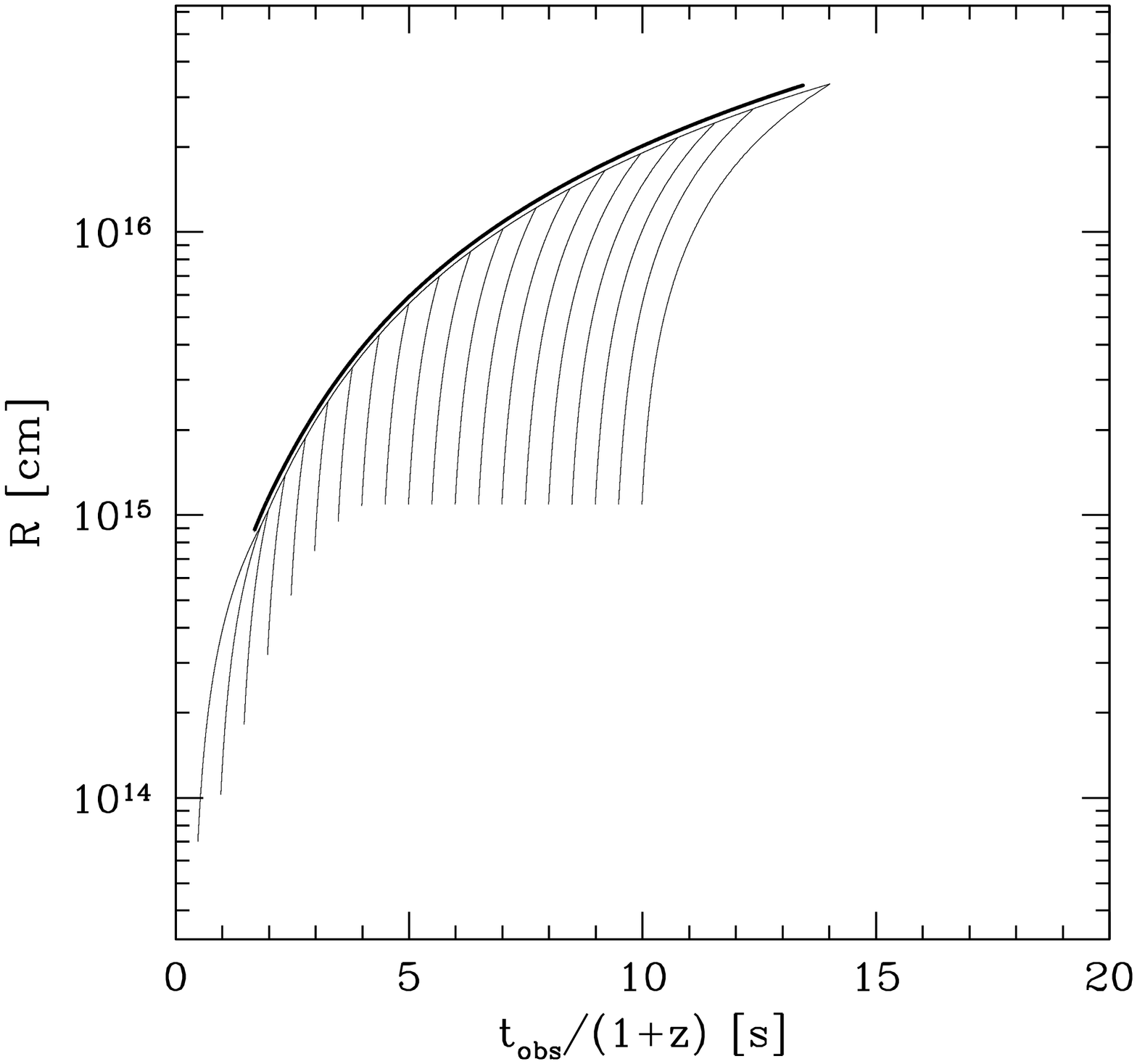} \\
\includegraphics[scale=0.275]{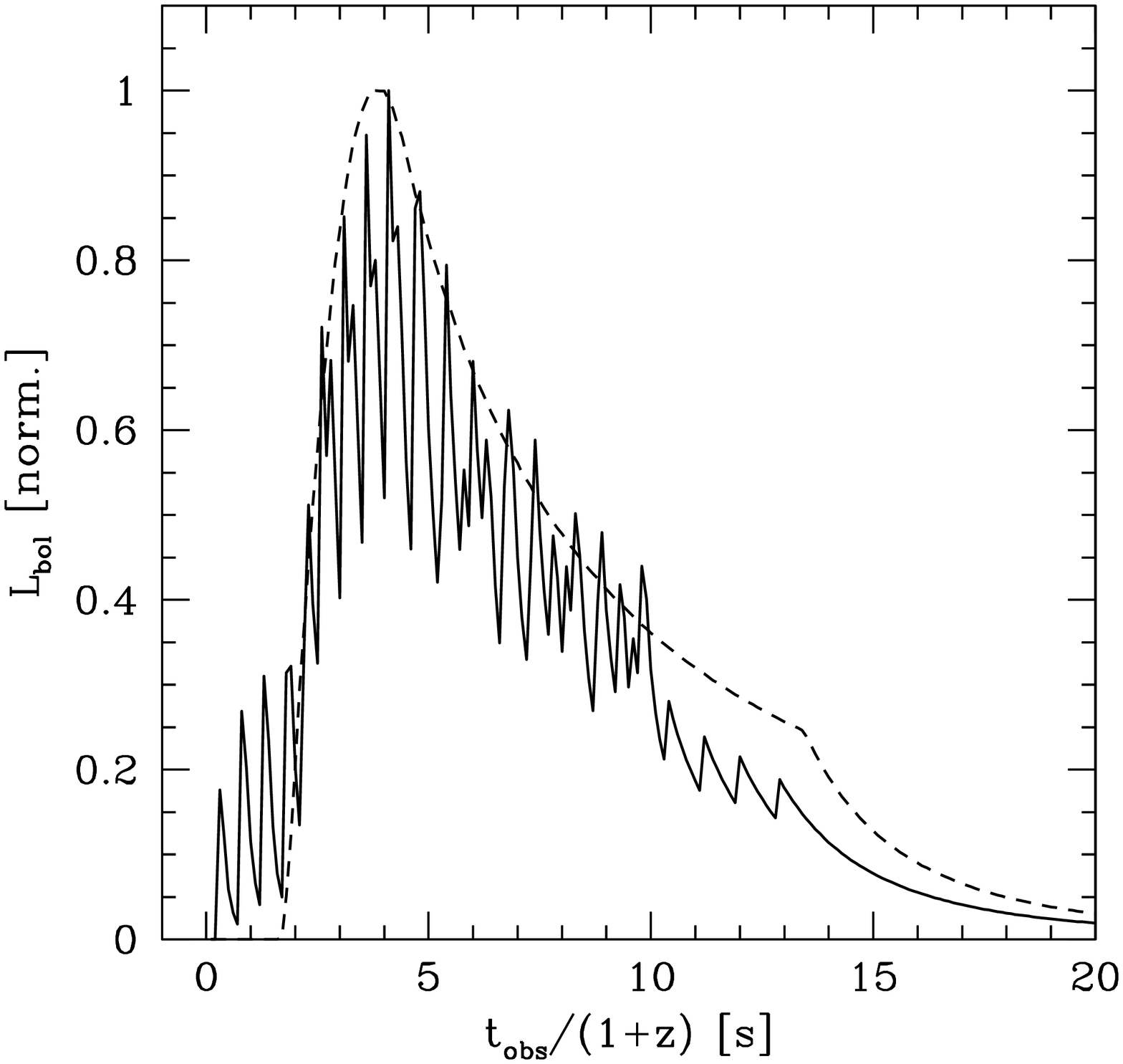} & \includegraphics[scale=0.275]{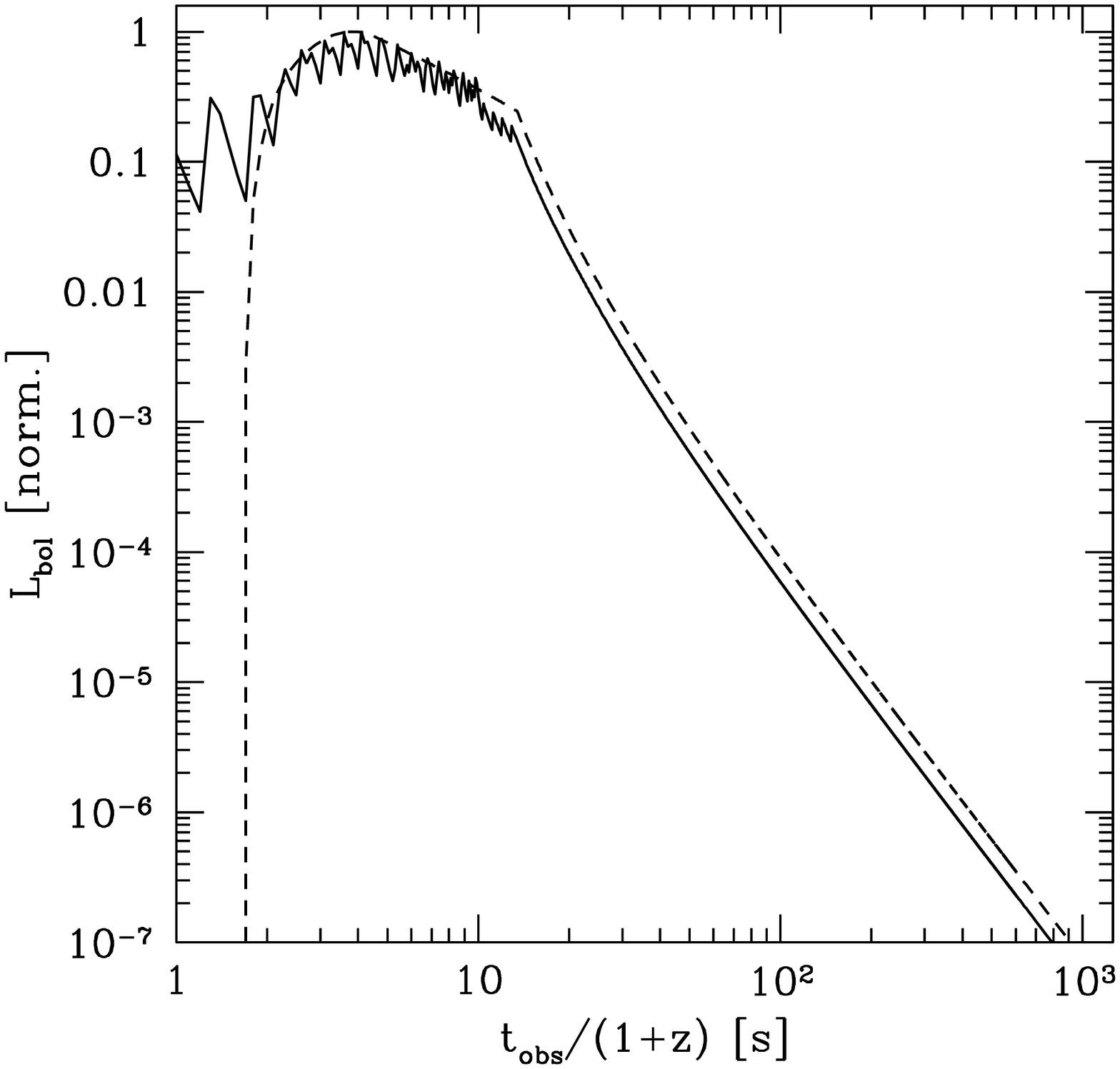} 
\end{tabular}
\end{center}
\vspace*{-5ex}

\caption{\textbf{Early steep decay from high latitude emission in the internal shock framework.} 
Two examples of synthetic GRBs with a smooth (dashed line) or a highly variable (solid line) light curve. 
\textit{Top left}: initial distribution of the Lorentz factor in the flow as a function of injection time $t_{inj}$
(a short timescale (0.5 s) fluctuation of the Lorentz factor is added to produce the variable burst). In the two cases\referee{,} the injected kinetic power is constant;
\textit{Top right}: shock radius (smooth light curve case: thick line; variable light curve case: thin line) as a function of observer time, showing that the maximum values reached by the emission radius, i.e. $R_{\gamma}$, 
are comparable in the two cases; 
\textit{Bottom}: bolometric light curve %in 
\referee{on either a} linear (left panel) or logarithmic (right panel) scale.
The high latitude contribution is
very similar for both the smooth and variable light curves.}
\label{fig_is_complex}
\end{figure*}

It has sometimes been argued (e.g. \citealt{lyutikov_2006, kumar_2007}) that the internal shock model 
(e.g. \citealt{rees_1994, kobayashi_1997, daigne_1998}) is 
%not compatible 
\referee{incompatible}
with the high latitude scenario. 
The underlying argument is based on the assumption that internal shocks take place at a typical radius $R_{\rm IS}$  
estimated from the shortest variability timescale $t_{\rm var,min}$ observed in the prompt $\gamma$-ray light curve. 
In this simplified context\referee{,} the inferred radius
%radius inferred for internal shocks 
$R_{\rm IS}\simeq 6 \ 10^{13}\,(\Gamma/100)^2\,(t_{\rm var,min}/0.1 \rm s) \ \rm cm$
is much too small 
%when 
\referee{compared} 
to the high latitude scenario constraints given by Eqs. \ref{eqn_contrainte1}-\ref{eqn_contrainte2}. 

However, observed $\gamma$-ray light curves cover a wide range of variability timescales going from the shortest one $t_{\rm var,min}$ 
to the longest one $t_\mathrm{var,max}$. \referee{The} power density spectra of GRB light curves show that $t_\mathrm{var,max}$ is much longer than $t_\mathrm{var,min}$ and close to the whole burst duration $t_\mathrm{burst}$ (see e.g. \citealt{beloborodov_2000, guidorzi_2012}). 
In the internal shock framework\referee{,} the variability timescales in the light curve reflect those of the 
relativistic outflow. 
%Then the 
\referee{The} initial Lorentz factor  must \referee{then} vary on timescales from $t_\mathrm{var,min}$ to $t_\mathrm{var,max}$ and
the radius of \referee{the} internal shocks extends from  
a minimum $R_{\rm IS,min} \simeq 2 \Gamma^2 c t_{\rm var,min}$ to a maximum 
$R_{\rm IS,max} \simeq 2 \Gamma^2 c t_{\rm var,max}$\referee{, which}
%that
 is very close to  
$R_*  \simeq 2 \Gamma^2 c t_{\rm burst}$. 
This implies that the the high latitude emission of the last internal shocks has a characteristic timescale 
$\tau_{\rm HLE} \simeq t_{\rm var,max} \la t_{\rm burst}$, leading to a decay index $\alpha\ga 3$, in good agreement with XRT data.
Internal shocks are therefore compatible with the requirements of the high latitude scenario
even in highly variable bursts. 

\begin{figure}
\begin{center}
\includegraphics[scale=0.28]{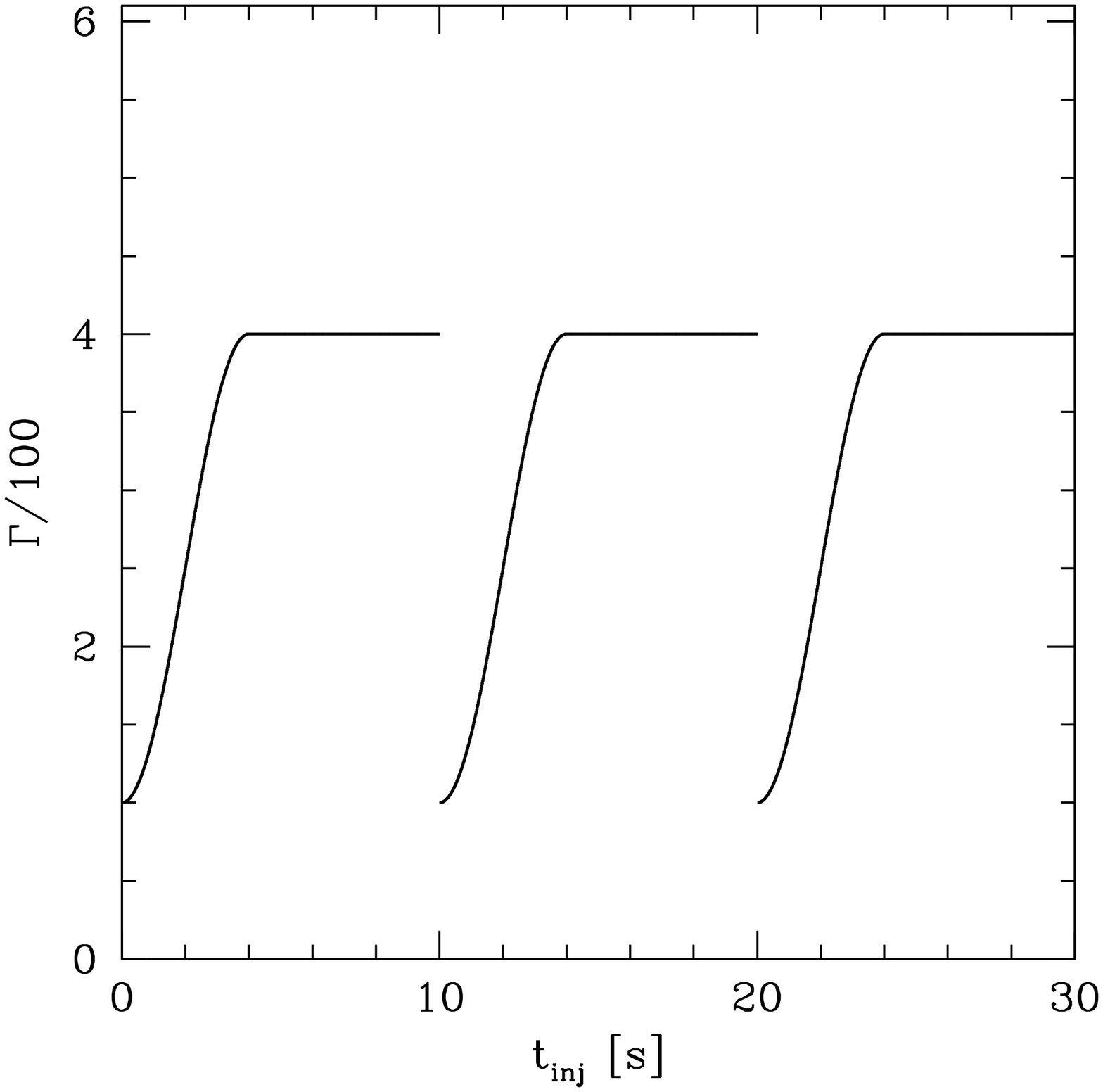}\\
\vspace*{-2ex}

\includegraphics[scale=0.28]{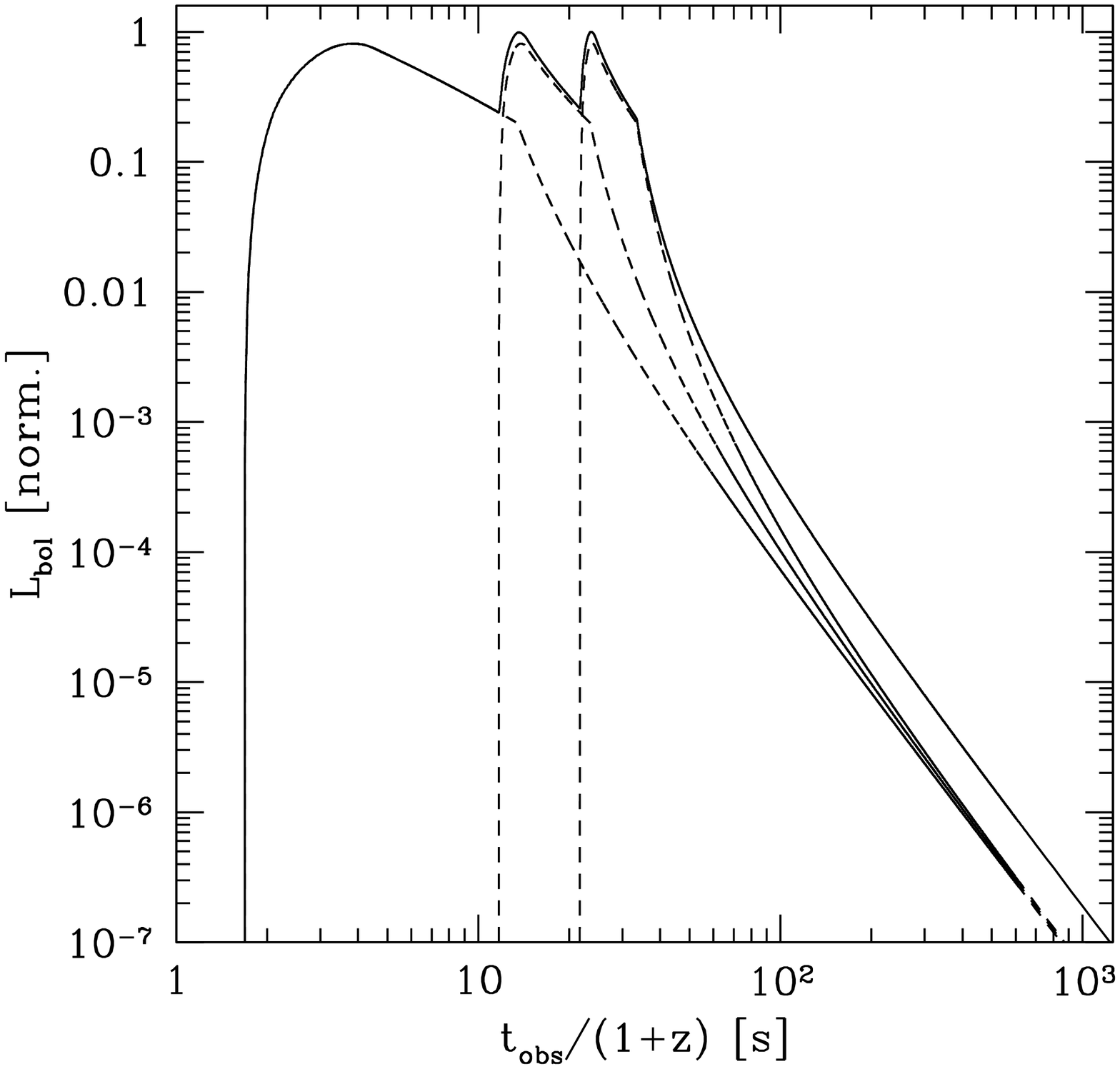} 
%\includegraphics[scale=0.3]{fig_2b.eps} 
%\begin{tabular}{cc}
%\includegraphics[scale=0.3]{fig_2a.eps} & \includegraphics[scale=0.3]{fig_2b.eps} 
%\end{tabular}
\end{center}
\vspace*{-5ex}

\caption{\textbf{Early steep decay from high latitude emission in the internal shock framework:  a multi-pulse burst.} \textit{Top:} initial distribution of the Lorentz factor; \textit{Bottom:} bolometric light curve 
%in
\referee{on a} logarithmic scale (solid line). The contribution of each individual pulse is plotted 
%in 
\referee{as a}
dashed line.
%Each of the three main regions in the initial outflow leads to a well separated pulse whose individual contribution is ploted in dashed line.
}
\label{fig_is_complex_3pulses}
\end{figure}

This result is illustrated %in
\referee{by} the synthetic GRBs shown in \fig{fig_is_complex} and~\ref{fig_is_complex_3pulses} where the dynamics of internal shocks has been computed with the multiple shell model of \citet{daigne_1998}. The  example in \fig{fig_is_complex} shows a mono-pulse  burst (bottom left panel), either with (solid line) or without (dashed line) sub-structure. As seen from the initial distribution of the Lorentz factor (top left panel), the maximum variability timescale $t_\mathrm{var,max}$ is the same in both cases but the 
%smallest
\referee{shortest} variability timescale $t_\mathrm{var,min}$ is very different. The plot showing the location of the shocks (top right panel) clearly illustrates how the final radius -- and therefore the high latitude emission -- is governed \referee{only} by $t_\mathrm{var,max}$. 
%  only. 
The decay of the flux at the end of the prompt phase (bottom right panel) is therefore very similar in both cases, with a steep decay index reaching $\alpha\simeq 3$ after a short and smooth transition. The example in \fig{fig_is_complex_3pulses} corresponds to a more complex light curve with three isolated pulses\referee{, where}
%. Here, 
$t_\mathrm{var,max}\simeq t_\mathrm{burst}/3$.  The high latitude emission of each pulse is plotted 
%in 
\referee{as a}
dashed line. 
%\referee{After addition of the three pulses,}
%They \referee{are superimposed} and  
\referee{The final decay index of the total light curve} is $\alpha\simeq 9$ before reaching $\alpha\simeq 3$ for $t_\mathrm{obs}\ga 2 t_\mathrm{burst}$ (see also \citealt{genet_2009}).
%after a short and smooth transition where $\alpha$ is steeper. 
We checked that adding substructure on \referee{shorter}
%smaller
 timescales does not affect this result.
As a last example, the `naked burst' GRB 050421 is a good case of a GRB with a variable light curve followed by an early steep decay 
\referee{that}
%which 
was well observed by the {XRT} 
%due to an
\referee{because of its}
 especially long duration \citep{godet_2006}. 
This burst 
%has been 
\referee{was}
modeled in \referee{detail} by \citet{hascoet_2011}. The prompt emission is associated 
%to 
\referee{with} internal shocks and
% and is computed using the same approach as in this letter. 
%It is found that
 the high latitude emission is in excellent agreement with {XRT} data (see Figure~2 in \citealt{hascoet_2011}).

%The observed short timescale variability in the synthetic light curves is associated to the initial short timescale variability in the Lorentz factor.
%As expected from the estimates above, as long as the initial variability in the outflow occurs on timescales extending to $t_{\rm var,max} \la t_{\rm burst}$ the final radius of the internal shock phase is large enough to produce a high latitude emission in agreement with observations.
%Indeed the high latitude emission of the last internal shocks have a characteristic timescale $\tau_{\rm HLE} \simeq t_{\rm var,max} \la t_{\rm burst}$.
%From Eq.~\ref{eqn_alpha} values of the initial temporal decay index of the early steep decay phase are predicted in the range $\alpha = 3 \rightarrow 9$,
%in agreement with observations (e.g. \citealt{beloborodov_2000, guidorzi_2012}), for $0.3 \la t_{\rm var,max} / t_{\rm burst} \la 1$.
%As power density spectra of observed prompt GRB light curves show that variability is indeed present up to a timescale $\sim t_{\rm burst}$,
\textit{We conclude from this analysis that the internal shock model naturally reproduces the {XRT} early steep decay phase in the high latitude scenario.}
%It should be noted
\referee{We note} that the value of $\alpha$ discussed above can be affected by two additional effects : 
{\it (i)} a spectral effect when the flux is integrated in a relatively narrow band (as 
%it is 
\referee{in}
the case 
%for
\referee{of}
 \textit{Swift}-{XRT}). 
%Then 
The expected temporal decay slope is \referee{then} $\alpha \simeq 2 + \beta$, 
where the spectral slope $\beta$ ($F_{\nu} \propto t^{-\alpha} \nu^{-\beta}$) can be larger than one; 
{\it (ii)} an emission that is already anisotropic in the comoving frame of the emitting material (e.g. \citealt{beloborodov_2011}). Finally, we %have 
\referee{limit our} 
%the 
discussion to the bolometric flux for simplicity. If the peak energy of the spectrum radiated by the last internal shocks is above the spectral range of the XRT (i.e. above 10 keV), the peak energy of the high latitude emission crosses the XRT range, leading to spectral evolution during the early steep decay (see e.g. \citealt{genet_2009}). 
%Such an 
\referee{This} evolution is seen in many XRT observations \citep{zhang_2007}.

%In some cases the observed slope of the early steep decay is steeper (with $\alpha = 3 \rightarrow 9$) than the value $\alpha_{bol} = 3$ 
%expected for a bolometric light curve. This could be due {\it (i)} to a spectral effect: when the flux is integrated 
%in a relatively narrow band (as it is the case for \textit{Swift}-{XRT}), the expected temporal decay slope is 
%$\alpha \simeq 2 + \beta$, where the spectral slope $\beta$ ($F_{\nu} \propto t^{-\alpha} \nu^{-\beta}$) can be 
%larger than one; {\it (ii)} 
%to an emission that is already anisotropic in the comoving frame of the emitting material (e.g. \citealt{beloborodov_2011}) or 
%finally {\it (iii)} to the fact that the decay \referee{timescale} $\tau_{\rm HLE}$ is moderately smaller than the burst duration $t_{\rm burst}$.
%This can be the case if the longest variability timescale in the outflow, $t_{\rm var, max}$, is slightly smaller than the total duration of the relativistic ejection, i.e. $t_{\rm HLE} \simeq t_{\rm var, max} \la t_{\rm burst}$.
%A typical example would be a burst made of $N$ successive pulses of similar duration $\sim t_{\rm burst}/N$
%(see e.g. \citealt{genet_2009}). 

\subsection{Photospheric emission}

In 
%the case of
 photospheric models\referee{,} the evolution of the high latitude flux is more complex than for a simple flashing sphere -- the last scattering 
region has a finite width and is not spherically symmetric (with the last scattering radius increasing with latitude). 
However\referee{,} a detailed calculation (e.g. \citealt{peer_2008,beloborodov_2011}) shows that the characteristic initial decay timescale remains 
$\tau_{\rm HLE} \simeq R_{\gamma} / 2\Gamma^{2} c$, where $R_{\gamma}$ is now the photospheric radius $R_{\rm ph}$ 
for on-axis photons. 
In the case of a ``classical'' non dissipative photosphere, $R_{\rm ph}$ is given by (e.g. \citealt{piran_1999, daigne_2002})
\begin{equation}
\label{eqn_rph}
R_{\rm ph} \simeq 6 \ 10^{12} \ (\dot{E}_{\rm iso} / 10^{52} \ \rm erg \ s^{-1}) \ (\Gamma/100)^{-3} \ \rm cm\ , 
\end{equation}
where $\dot{E}_{\rm iso}$ is the isotropic equivalent kinetic power in the outflow, leading to 
\begin{equation}
\tau_{\rm HLE} \simeq 10^{-2} \ (\dot{E}_{\rm iso} / 10^{52} \ \rm erg \ s^{-1}) \ (\Gamma/100)^{-5} \ \rm s\ .
\end{equation}
%
%It appears that f
For standard values of the parameters $\tau_{\rm HLE}\ll t_{\rm burst}$\referee{,} so that 
the early steep decay cannot be explained by high latitude emission. Reducing the Lorentz factor at the photosphere to $\Gamma \la 20$
can increase $R_{\rm ph}$ to an acceptable value but one is then confronted \referee{with}
%to
a drop 
%of 
\referee{in} the photospheric luminosity even before the high latitude emission sets in (since $L_{\rm ph}\propto \Gamma^{8/3}$, e.g. \citealt{meszaros_2000,daigne_2002}). 
The same conclusion holds 
(as $R_{\rm ph}$ remains very close to the value given by Eq. \ref{eqn_rph}) 

in models where a dissipative 
process is invoked at the
photosphere to transform the seed thermal spectrum into a \referee{non-thermal} Band spectrum 
(see e.g. \citealt{rees_2005, beloborodov_2010}).

\textit{In photospheric models, the early steep decay must therefore be directly produced by a declining activity of the central engine. }
This implies that a physical mechanism, common to most GRBs, 
governs the late activity in a generic way, in contrast 
%with
\referee{to}
 the diversity of the prompt gamma-ray light curves, 
which also reflect the activity of the central engine.

\subsection{Magnetic reconnection}

\referee{Gamma-ray burst} models where the prompt emission comes from magnetic reconnection are still at an early stage of development and do not offer the same level of prediction as the two other families of models discussed above. 

In a first situation investigated by \referee{\citet{drenkhahn_2002}, \citet{drenkhahn_2002b}, and \citet{giannios_2008},} the prompt emission is produced by a gradual 
reconnection process that begins below the photosphere and extends above. 
The dissipation process typically ends at a radius $R \simeq 10^{13}$ cm, which remains too small for the high latitude scenario. 
In the model proposed by \citet{mckinney_2012}, reconnection remains inefficient below the photosphere before it enters 
a rapid collisionless mode at $R_{\rm diss}\sim 10^{13} - 10^{14}$ cm and \referee{catastrophically dissipates}  the magnetic energy of the jet.
This radius range becomes compatible with the high latitude scenario for GRBs of duration $t_{\rm burst} \simeq 1$s 
but is still too small for most %of the
 long GRBs.
 % population. 
 \referee{ However, there are remaining theoretical} uncertainties 
 %regarding 
 \referee{about} the dissipation rate, 
%subsist in these models, 
and 
%one cannot exclude that 
future calculations 
%will
\referee{may} predict a larger radius 
for the end of the reconnection process.

A full electromagnetic model was proposed by \citet{lyutikov_2003} where the energy is dissipated at large radii ($\ga$~$10^{16}$ cm)
when the interaction of an electromagnetic bubble with the shocked external medium becomes significant and current instabilities 
develop at the contact discontinuity.
In this model\referee{,} the gamma-ray variability is generated by the emission of ``fundamental emitters'' that are driven (by the dissipative process) 
into relativistic motion with random directions (some kind of relativistic turbulence) in the \referee{frame of the} main outflow\referee{, which is}
 %frame\referee{, where the}
% -- main outflow that is 
also highly relativistic. Different authors have tentatively computed the expected bolometric light curves: in the examples 
presented by \referee{\citet{lyutikov_2006}, \citet{lazar_2009}, and \citet{narayan_2009},} the high latitude contribution smoothly connects to the end of a highly variable prompt phase. These results however require some degree of adjustment between different dynamical parameters. It is \textit{a priori} 
decided that $R_{\gamma} \simeq \Gamma^2 c t_{\rm burst}$ -- which is not an intrinsic consequence of the model, 
%contrary
\referee{in contrast} to what happens with
internal shocks.

Finally\referee{,} in the model of \citet{zhang_2011}, a self-sustained magnetic process is triggered by prior internal shocks that are 
inefficient (because of a high degree of magnetization) but \referee{leave}
%get 
the magnetic field lines entangled. Some ``dynamical memory''
of the shocks is conserved and the conclusions of \sect{subsect_internal_shocks} basically apply. However\referee{,} the details of the
physical mechanisms 
involved in this model remain to be clarified and quantified. 

\section{Conclusion}
\label{section_conclusion}

The results of this letter emphasize the importance of the steep decay phase revealed by \textit{Swift} observations
of the early X-ray afterglow.
An attractive way 
%to explain 
\referee{of explaining}
this phase is to suppose that it is produced 
by the high latitude emission of the last contributing shells. We have shown that this assumption leads to strong constraints on the
radius $R_{\gamma}$ where the prompt emission ends. In particular, the associated \referee{timescale}, 
$\tau_{\rm HLE}\sim R_{\gamma}/2c\,\Gamma^2$ must be comparable to the burst duration $t_{\rm burst}$ to guarantee that the high latitude contribution is correctly
connected to the end of the prompt light curve.  
We have then checked 
%if 
\referee{whether}
these constraints are satisfied by different models
for the prompt phase, namely internal shocks, \referee{Comptonized} photospheric emission\referee{,} and magnetic reconnection.
\begin{itemize}
\item Internal shocks naturally fulfill the condition $\tau_{\rm HLE}\sim t_{\rm burst}$ even in highly variable bursts, as the radius of the last shocks is governed by the longest variability timescale\referee{,} which is 
%of 
\referee{on} the order of $t_\mathrm{burst}$.
%This is because the geometrical timescale $\tau_{\rm HLE}$ is
%also the dynamical timescale for shell collision in the internal shock framework. 
\item Conversely\referee{,} in photospheric models $\tau_{\rm HLE}\ll t_{\rm burst}$, since the photospheric radius is typically 
several orders of magnitude smaller than the radius of internal shocks. 
The only way to produce the observed decay is then to suppose that it corresponds to an effective behavior of the central engine,
 which moreover should be common to most GRBs. 
\item In magnetic reconnection models, satisfying the constraints on $R_{\gamma}$ might be possible but is not naturally expected. 
It still requires the ad hoc assumption that $R_{\gamma} \simeq \Gamma^2 c t_{\rm burst}$,  which will have to be justified when a better description of the reconnection process becomes available.
\end{itemize} 

\vspace*{-2ex}

\begin{acknowledgements} 
The authors acknowledge the French Space Agency (CNES) for financial support.
R.H.'s PhD work is funded by a Fondation CFM-JP Aguilar grant.
\end{acknowledgements}
\vspace*{-5ex}

\bibliographystyle{aa}
\bibliography{main}

\end{document}